\documentclass[12pt]{article} 
\usepackage{epsfig} 
\def\bea {\begin{eqnarray}}
\def\eea {\end{eqnarray}}
\def\be{\begin{equation}}
\def\ee{\end{equation}}

\begin{document}
\begin{titlepage} 
\begin{flushright}
ADP-09-16/T694 \\
JLAB-THY-09-1097
\end{flushright} 
\begin{center}
\vspace{5mm}
\Large{\bf Understanding the proton's spin structure  }
\\
\vskip 0.5cm  
{\large F. Myhrer$^{a}$ and A.W. Thomas$^{b,c}$}
\vskip 0.5cm 
$^{a}$~{\large Department of Physics and Astronomy,
University of South Carolina, 
Columbia, SC 29208 }\\
$^{b}$~{\large Jefferson Lab, 12000 Jefferson Ave.,
 \\ Newport News, VA 23606}\\
$^{c}$~{\large Centre for the Subatomic Structure of Matter and 
School of Chemistry and Physics, \\ The University of Adelaide, 
Adelaide SA 5005, Australia}
\end{center}
\centerline{(\today) }
\vskip 1cm
\begin{abstract}
We discuss the tremendous progress that has been towards an understanding 
of how the spin of the proton is distributed on its quark and gluon
constituents. This is a problem that began in earnest twenty years ago 
with the discovery of the proton ``spin crisis'' by the 
European Muon Collaboration. The discoveries prompted by that original 
work have given us unprecedented insight into the amount of spin 
carried by polarized gluons and the orbital angular momentum of the 
quarks.  

\end{abstract}
\end{titlepage}
\newpage 

\begin{center}
{\bf 1. Introduction} 
\end{center}

The spin structure of the proton, or how the spins of the quarks and  
polarized gluons
and their orbital angular momenta make up the total angular momentum 1/2, 
has been a puzzle since the discovery by the 
European Muon Collaboration (EMC) that 
the quarks appeared to contribute very little to the 
proton spin~\cite{Ashman:1987hv}. 
This surprising experimental result appears to be 
in contrast to the apparently successful, naive 
quark model descriptions of 
the structure of the proton, 
such as the 
proton's charge radius, magnetic moment and axial charge.  
The explanation of these nucleon structure observables has its basis in 
%the  
QCD, the theory of strong interaction, which is invariant under 
chiral transformations provided that the $u$ and $d$ quarks are massless. 
Furthermore, chiral perturbation theory (ChPT), 
an effective low-energy theory of QCD, 
assumes that the Goldstone Boson of the theory is the effective pion-field, 
which more properly should be considered a coherent 
state of quarks and anti-quarks. 
We know that chiral symmetry is explicitly broken by 
non-zero quark masses and according to ChPT the pion mass 
is generated by the small $u$ and $d$ quark masses $m_q$, leading  
to the relation  $m_\pi^2 \propto m_q$. 

These ideas have established 
that the low-momentum structure of the nucleon 
consists of a core of three confined, 
interacting valence quarks (the ``bare" nucleon) plus a pion cloud 
which interacts with the valence quarks of the quark core 
in a manner dictated by chiral symmetry. 
This description of the nucleon was however not capable of reproducing 
the {\it initial} EMC measurement~\cite{Ashman:1987hv} of the 
very small quark spin contribution to the proton spin. 
As will be presented in this topical review 
the above description of the proton's structure does explain the 
updated measurements of the quark spin contribution.

%\vspace{5mm} 

The proton spin structure is experimentally  
explored in deep inelastic scattering (DIS) 
and polarized proton proton reactions   
where the four-momentum transfer squared, $Q^2$, is large compared to 
the hadronic scale $\sim 1$(GeV/c)$^2$. 
At these large values of $Q^2$ it appears natural to use simple parton model 
considerations to analyze the experimentally measured structure functions 
which are functions of $Q^2$ and the Bjorken variable $x$. 
The measured dependence of the structure functions on $Q^2$ is 
consistent with the expected perturbative QCD behavior 
down to surprisingly low $Q^2$ values. 
However, we know that even at high $Q^2$ the three confined valence 
quarks in the proton  
retain their transverse degrees of freedom, 
such as the transverse momentum $k_\perp$, 
and these transverse degrees of freedom can have a 
non-negligible influence on 
the interpretation of the experimental spin-dependent data.  
We will elaborate on this topic in the fifth section of this review. 

%\vspace{5mm}

In the next section we
briefly present experimental data which are 
relevant to the proton spin structure.  
The focus will be on recent experimental results which 
appear to indicate that polarized gluons contribute little to the 
proton spin. 
In the third section we concentrate on 
well-known non-perturbative aspects of QCD 
which successfully explain 
the recently determined first moment value, $\Sigma$, of the measured 
proton spin structure function $g_1(x)$. 
In the fourth section we discuss the recent lattice evaluations of 
the $u$- and $d$- quark angular momenta.
Finally, before summarizing this review, 
we will discuss  
the transverse momentum distributions of the quarks which affect the  
spin distributions of the proton.  
This fifth section also include a brief discussion of 
the transverse structure functions 
measurements necessary in 
order to verify experimentally the consequences implied by our 
explanation of $\Sigma$, namely 
that the quark and antiquark orbital angular momentum contributions 
to the proton spin are sizable.

%\vspace{5mm}

%%%%%%%%%%%%%%%%%%%%%%%%%%%%%%%%%%%%%%%%%%%%%%%%%%%%%%%%%%%%%%%%%%%%%%%%%
\begin{center}
{\bf 2. A summary of the experimental data }
\end{center}
%%%%%%%%%%%%%%%%%%%%%%%%%%%%%%%%%%%%%%%%%%%%%%%%%%%%%%%%%%%%%%%%%%%%%%%%%

The EMC experimental result that $\Sigma$ is small~\cite{Ashman:1987hv}, 
the so-called proton ``spin crisis", 
generated a tremendous effort in order to map out the quark and gluon 
distributions of energy, momentum, spin and angular momentum 
of the proton. 
The experimental effort at 
CERN~\cite{Ashman:1987hv,SMC,Ageev:2005gh,Alexakhin:2007}, 
DESY~\cite{Airapetian:2007mh}, JLab~\cite{JLAB}, 
RHIC~\cite{Abelev:2006uq,PHENIX} and SLAC~\cite{SLAC} over 
the past two decades has been very
impressive and today several crucial pieces 
of information have been established. 
The longitudinal spin structure function $g_1(x, Q^2)$ 
has been measured over a range of  
values of $Q^2$ and for $x$ down to 10$^{-4}$ -- although 
at the lower x values 
%end 
the COMPASS data~\cite{Alexakhin:2007} on deuterium is 
%well 
below $Q^2 = 1$ GeV$^2$.

The quark spin contribution, $\Sigma$, measures the quark 
(and antiquark) helicity
along the longitudinal proton spin minus the 
quark (and antiquark) helicity antiparallel to 
the proton spin. 
At very large $Q^2$ values 
the integral of the proton spin structure function, $g_1(x,Q^2)$,  
can be written as 
\begin{eqnarray}
\Gamma^p(Q^2) &=& \int_0^1 {\rm d}x \; g_1^p(x,Q^2) = 
\frac{ c_1(Q^2)}{12} \left[ g_A^{(3)} + \frac{1}{3} \  g_A^{(8)} \right] 
+\frac{c_2(Q^2)}{9} \; \Sigma 
\label{eq:Bjorken1}
\end{eqnarray}
The low-energy nucleon axial coupling constant, $g_A^{(3)} = g_A$ 
is very well known from neutron $\beta$-decay, while $g_A^{(8)} $ 
has been determined from semi-leptonic hyperon decay with an error 
that is often quoted as 5\% but may be as large as 20\% \cite{Jaffe:1989jz}.  
The flavor-singlet axial-current matrix element, $\Sigma$, 
is the focus of this review. 
%
%In section three we will present  
The non-singlet and singlet ``radiative" 
coefficients, $c_1(Q^2)$ and
$c_2(Q^2)$ respectively, 
%which 
have been evaluated in QCD perturbation theory 
at the three-loop level~\cite{Larin:1991tj,Larin:1994va}. 
As  $Q^2 \to \infty$ both coefficients approach 1. 
%%%%%%%%%%%%%%%%%%%%%%%%%%%%%%%%%%%%%%%%%%%%%%%%%%%%%%%%%%%%%%%%%%%%%%%%

A summary of the status and recent experimental results on 
the spin structure of the nucleon 
can be found in Ref.~\cite{Kuhn:2009}.
Additionally, we refer to 
the recent work of Bass~\cite{Bass:2009} who 
stresses that the isovector and 
isoscalar components of the measured 
spin structure function $g_1(x)$ behave very differently at small
$x$-values~\cite{Airapetian:2007mh}, 
and he presents possible reasons for these different behaviors. 
%%%%%%%%%%%%%%%%%%%%%%%%%%%%%%%%%%%%%%%%%%%%%%%%%%%%%%%%%%%%%%%%%%%%%%%

%\vspace{5mm} 

Unlike the early EMC result that the quark spin contribution $\Sigma$ was 
consistent with zero, $14 \pm 9 \pm 21\% $~\cite{Ashman:1987hv}, 
today we know that the sum of the helicities of the quarks 
in the proton is about a third of the total 
spin~\cite{Ageev:2005gh,Alexakhin:2007,Airapetian:2007mh},  
\begin{eqnarray}
\Sigma =  0.33 \pm 0.03 
({\rm stat.}) \pm 0.05 ({\rm syst.}) \; . 
\label{eq:SigmaEXP}
\end{eqnarray} 
This result is small compared to  $\Sigma_{quark} \sim $ 0.67, 
the expected value obtained when one considers 
the relativistic motion of the confined valence quarks, 
which we will discuss in the next section.  

%\vspace{5mm}

The {\it initial}, extremely low EMC value of $\Sigma$ raised the
exciting possibility that  the proton could contain 
a substantial quantity of polarized gluons, which can contribute
to $g_1^p(x)$ through 
the axial $U(1)$ anomaly~\cite{Efremov:1988zh,
Altarelli:1988nr,ccm88,la88,Bodwin:1989nz,Bass:1991yx}. 
The essential point is that the flavor-singlet axial current 
is not conserved because of this anomaly. To account for the 
effect of the anomaly the matrix element of the axial charge $\Sigma$ 
in Eq.(\ref{eq:Bjorken1}) could be written as
(see the review by Bass~\cite{Bass:2004xa}):
% in e.g. Ref.~\cite{ccm88}: 
%
\begin{eqnarray}
\Sigma = \Sigma_{quark} - \ \frac{N_f \alpha_s(Q^2)}{2\pi}\ 
\Delta G(Q^2) \, . 
\label{eq:UA1}
\end{eqnarray}
Here $\Sigma_{quark}$ is the quark model prediction and the 
second term is the contribution of polarized gluons arising from the 
axial anomaly. 
In the limit $Q^2 \to \infty$ the product 
$\alpha_s(Q^2) \ \Delta G(Q^2)$ has a non-zero value --  
see e.g. Refs.~\cite{Efremov:1988zh,Altarelli:1988nr,ccm88}. 
The question pursued by several experimental groups is whether 
the polarized gluon content of the proton, $\Delta G(Q^2)$, 
is large enough to explain the measured value of $\Sigma$ 
reported in Eq.(\ref{eq:SigmaEXP}). 

%\vspace{5mm} 

If we ascribe the unexpectedly small observed $\Sigma$ 
value in Eq.(\ref{eq:SigmaEXP})
as being entirely caused by the 
polarized gluons in the proton, we can estimate the value of 
$\Delta G(Q^2)$ expected from experiments. 
At a scale of $Q^2 \simeq 3$ (GeV/c)$^2$ we know $\alpha_s(Q^2) \simeq 0.3$. 
When we include  only the 
relativistic corrections due to the confined quarks' motions 
in the proton giving $\Sigma_{quark} \simeq 0.67$, 
we obtain from Eqs.(\ref{eq:SigmaEXP} and (\ref{eq:UA1}) with three flavors
a value for $\Delta G(Q^2) \simeq 2.4$. 
The recent experimental data however indicate that 
the polarization of the gluons is much smaller, typically 
only one tenth of this.
%\footnote{ 
%%%%%%%%%%%%%%%%%%%%%%%%%%%%%%%%%%%%%%%%%%%%%%%%%%%%%%%%%%%%%%%%%%%%%%%%% 
%
%INCLUDE HERE ADDITIONAL DISCUSSION AND SOME REFERENCES FROM 
%SPIN-2008 CONFERENCE, OCTOBER 2008: 
%http://faculty.virginia.edu/spin2008/ %scientific_program.html 
%and click on {\it program}
%
%%%%%%%%%%%%%%%%%%%%%%%%%%%%%%%%%%%%%%%%%%%%%%%%%%%%%%%%%%%%%%%%%%%%%%%%%% 
% } 
%  

At the SPIN2008 conference Rondio 
summarized the status of the polarized gluon 
experiments~\cite{Rondio08} and concluded that
most likely the gluon polarization is small.
Most recently the COMPASS Collaboration 
used an idea proposed in Ref.~\cite{ccm88} 
to measure the polarized gluons in the proton.
They scattered polarized muons off a 
longitudinal polarized deuteron target and detected  
charm mesons which are  presumed to originate from the 
sub-process $\gamma + g \to q \bar{q}$ and are produced at high $Q^2$. 
Their result, namely that 
$\Delta g(x)/g(x)$\footnote{
The Bjorken $x$ ranges from 10$^{-5}$ to 0.6 in 
Ref.~\cite{Alekseev:2009}. 
} 
is negative 
for $<x> \simeq 0.11$~\cite{Alekseev:2009},   
led to their conclusion that 
``This is a hint for a small value of the 
first moment, $\Delta G$, of the gluon helicity distribution, 
although this in principle does not exclude a large value." 
The measurements reported a year earlier at Pacific-SPIN07 of 
inclusive $\pi^0$ 
jets at RHIC are best fit with $\Delta G$ 
consistent with zero~\cite{Jacobs-Pacific,Abelev:2006uq}.  
Bianchi~\cite{Bianchi,Airapetian:2007mh}  
reported a small but non-zero $\Delta G /G \sim 0.08$ at Pacific-SPIN07 -- see  
 also the presentation by Kabuss~\cite{Kabuss_Pacific} at the same conference. 
%%%%%%%%%%%%%%%%%%%%%%%   NEW in 003 version   %%%%%%%%%%%%%%%%%%%%%%%%%%%%%%%

A recent global analysis of parton helicity densities by 
deFlorian {\it et al.}~\cite{deFlorian:2008} 
incorporated DIS data as well as the newly published $A_{LL}$ measurements at 
RHIC~\cite{Adare:2009}. 
Based on this analysis Ref.~\cite{deFlorian:2008} concluded that 
$\Delta G$ is small.
However, $\Delta g(x)$ has only been measured in a limited $x$-interval, e.g.,  
$0.06 < x < 0.4$ in the latest RHIC measurements~\cite{Adare:2009}, and 
it is desirable to expand the measurements to a larger $x$-range before a firm
conclusion on a precise value $\Delta G$ can be reached
as stressed in the talks by Aidala~\cite{Aidala:2009} and   
Ellinghaus~\cite{Ellinghaus:2009} 
at the SPIN2008 conference. 
%%%%%%%%%%%%%%%%%%%%%%%%%%%%%%%%%%%%%%%%%%%%%%%%%%%%%%%%%%%%%%%%%%%%%
%

It should be noted that in recent preprints~\cite{lss08} Leader {\it et al.} 
have some critical comments on the extraction of $\Delta G$ from the 
earlier COMPASS Collaboration data~\cite{Alexakhin:2007}. 
Leader {\it et al.} find that especially the CLAS data demand 
higher twist (HT) terms and they state that
``They [COMPASS] do {\it not} include HT terms". 
Ref.~\cite{lss08} continues to say: 
``We fail to find negative $\Delta g(x)$ fits {\it without} HT terms!". 
In their presentation at SPIN2008 they~\cite{lss08} stated that 
present day data cannot distinguish between a 
positive or negative value for $\Delta G$. 
%\footnote{ 
%%%%%%%%%%%%%%%%%%%%%%%%%%%%%%%%%%%%%%%%%%%%%%%%%%%%%%%%%%%%%%%%%%%%%%%%%%%%%%%% 
% {\bf [I (FM) will search the SPIN2008 site for more comments on these points]}
%%%%%%%%%%%%%%%%%%%%%%%%%%%%%%%%%%%%%%%%%%%%%%%%%%%%%%%%%%%%%%%%%%%%%%%%%%%%%%%%
% } 
%
%
%%%%%%%%%%%%%%%%%%%%%%%%%%%%%%%%%%%%%%%%%%%%%%%%%%%%%%%%%%%%%%%%%%%%%%%%% 

If as a rough estimate we assume $\Delta G \simeq 0.2$ and 
$\alpha_s (Q^2) \simeq 0.3$ for $Q^2 \simeq 3$ (GeV/c)$^2$
we find for $N_f =3$ that the second term in Eq.(\ref{eq:UA1}) contributes 
about $0.03 \ll \Sigma_{quark}$. In other words, a value for 
$\Delta G$ much larger than 0.2 is necessary in order 
to reconcile the  
observed quark spin content  
of the proton with the expectations in a relativistic quark picture. 
%%%%%%%%%%%%%%%%%%%%%%%%%%%%%%%%%%%%%%%%%%%%%%%%%%%%%%%%%%%%%%%%%%%%%%%%%%

%\vspace{5mm}

%%%%%%%%%%%%%%%%%%%%%%%%%%%%%%%%%%%%%%%%%%%%%%%%%%%%%%%%%%%%%%%%%%%%%%%%%
\begin{center}
{\bf 3. The modern explanation of the value of $\Sigma$ }
\end{center}
%%%%%%%%%%%%%%%%%%%%%%%%%%%%%%%%%%%%%%%%%%%%%%%%%%%%%%%%%%%%%%%%%%%%%%%%%

There were two recent developments which inspired us to re-examine 
the evaluation of $\Sigma$ within the successful 
quark model description of proton structure.
First, as we already discussed, the current experimental evidence 
shows that polarized gluons cannot explain a major 
part of the observed reduction in $\Sigma$. 
Second, new studies in lattice QCD evaluations 
of the masses of the nucleon and $\Delta$ as a function of quark mass 
have resulted in the discovery~\cite{Young:2002cj,Young:2002rx} 
that the pion loops yield only $40 \pm 20$ MeV of the nucleon - $\Delta$ 
mass difference.
These two developments led us to 
reconsider an explanation for the 
$\Sigma$ value made 
shortly after the original EMC results were   
known~\cite{Myhrer:1988ap,Schreiber:1988uw,Hogaasen:1988,kkty89}. 

In this section we will demonstrate that  
well known, non-perturbative QCD aspects of nucleon structure, 
involving its pion cloud and the
quark hyperfine interaction mediated by an effective one-gluon exchange 
force combined with the relativistic motion of the confined quarks, 
not only explain the baryon magnetic moments and their semi-leptonic decays 
but also give a very satisfactory explanation of 
the modern experimental value of $\Sigma$~\cite{mt08}. 
A consequence of this new insight is that the missing spin should be 
accounted for by the orbital angular momentum of the quarks and anti-quarks, 
a topic we will discuss further in sections four and five. 

%\vspace{5mm}

In the limit  $Q^2 \to \infty$ 
the Bjorken sum rule, as derived from dispersion theory~\cite{Bjorken66}, 
says that the integral over the proton 
spin structure function, $g_1^p(x)$, equals a low-energy 
axial current matrix element of the proton with $s_z = + 1/2$; 
\begin{eqnarray}
\Gamma^p_1 = 
\int_0^1 {\rm d}x \; g_1^p(x, Q^2 \to \infty) 
= \langle p \uparrow | \sum_i \bar{\Psi}_i \; q_i^2 \gamma_5 \gamma_3 \; \Psi_i \; 
|\;  p \uparrow \rangle \; , 
\label{eq:Gamma}
\end{eqnarray}
where $q_i$ is the charge of quark $i$ (in units of the proton charge).  
The r.h.s. of Eq.~(\ref{eq:Gamma}) 
was written in Eq.~(\ref{eq:Bjorken1}) as 
the sum of an isovector, an $SU_F(3)$ octet and a flavor singlet component. 

In the naive parton model only the one-body axial currents are  
considered when the matrix element in Eq.(\ref{eq:Gamma}) is evaluated.
However,  the proton contains three confined, interacting valence quarks and 
below we will present the major
contributions to the axial current of 
the low-energy proton matrix element on the 
r.h.s. of Eq.~(\ref{eq:Gamma}). 
%%%
Specifically,  the matrix element of the axial current in Eq.(\ref{eq:Gamma}) 
includes contributions generated by 
two-quark axial operators~\cite{Myhrer:1988ap} 
and is strongly influenced by the pion cloud~\cite{Schreiber:1988uw}. 
In the following explanation of the measured value of $\Sigma$ 
we will use as guidance the results which are obtained by the 
cloudy bag model~\cite{Theberge:1980ye,Thomas:1982kv}, 
a model which successfully describes nucleon observables including the 
axial coupling $g_A$ -- e.g. Ref~\cite{Tsushima:1988}.

%%%%%%%%%%%%%%%%%%%%%%%%%%%%%%%%%%%%%%%%%%%%%%%%%%%%%%%%%%%%%%%%%%
\begin{center}
{\bf 3.1 Relativistic valence quark} 
\end{center}
%%%%%%%%%%%%%%%%%%%%%%%%%%%%%%%%%%%%%%%%%%%%%%%%%%%%%%%%%%%%%%%%%%%

Even at the time when the results of the EMC experiment were 
published it was known that
the motion of the confined quarks would reduce the value of $\Sigma$.  
The current $u$ and $d$ quark masses are small compared to the QCD scale,  
$\Lambda_{QCD} \simeq 300$ Mev/c, and in a 
space of dimension of 1 fm the light quark moves relativistically.  
A spin-up quark in an s-state has a lower p-wave Dirac component. For a 
proton with spin up this 
lower component naturally has spin-down and thereby reduces
the ``spin content" of the valence quark. 
In the bag model, which is a spherical confining cavity of radius $R = 1$ fm,  
it is an excellent approximation to work with massless $u$ and $d$ quarks
which have a 
minimal energy $E_q = \Omega/R$, where $\Omega \simeq 2.04$. 
The reduction factor for the axial charge of the light quarks in the bag,
compared with the non-relatistic limit, is 
$B= \Omega/3(\Omega - 1) \simeq 0.65$.  
This value changes very little if we use typical 
light quark current masses, 
$m_u \simeq 7$ MeV and $m_d \simeq 15$ MeV. 
We also note that even in modern relativistic models, where quark confinement 
is simulated by forbidding on-shell propagation 
through proper time regularization, the 
reduction factor is very similar. 
{}For example, Ref.~\cite{Cloet:2007em} finds a factor 0.67 and  
a similar result is found in Ref.~\cite{QCW98}. 
In other words, the valence quarks' orbital motion account for roughly 35\% 
of the nucleon spin. 

%\vspace{5mm}

%%%%%%%%%%%%%%%%%%%%%%%%%%%%%%%%%%%%%%%%%%%%%%%%%%%%%%%%%%%%%%%%%%
\begin{center}
{\bf 3.2 The quark-quark hyperfine interaction } 
\end{center}
%%%%%%%%%%%%%%%%%%%%%%%%%%%%%%%%%%%%%%%%%%%%%%%%%%%%%%%%%%%%%%%%%%%
%
\begin{figure}
\begin{center}
\epsfig{file=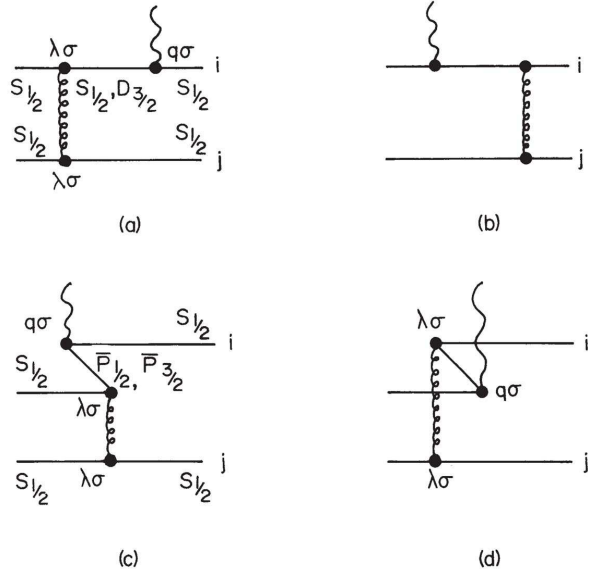, width=8cm}
\end{center}
\caption[]{We illustrate the quark-quark hyperfine 
contributions which involve an excited intermediate quark state. 
In the figures the external probe (top vertical wavy line) 
couples to the i'th quark which interacts with the second j'th quark via 
the effective one gluon exchange. The intermediate quark propagator is 
evaluated as a sum over confined quark modes. In Figs. (a) and (b) 
we illustrate the  three-quark intermediate states,  and in (c) and (d) 
the  one anti-quark and four quarks intermediate states. 
The mode sum converges rapidly and the lowest 
anti-quark $P_{1/2}$ and $P_{3/2}$ modes dominate 
the mode-sum~\cite{Hogaasen:1988jd}.}
\label{fig:I}
\end{figure}        
%%%%%%%%%%%%%%%%%%%%%%%%%%%%%%%%%%%%%%%%%%%%%%%%%%%%%%%%%%%%%%%%%%%%%%%%%%%%%%%%% 
%%%%%%%%%%%%%%%%%%%%%%%%%%%%%%%%%%%%%%%%%%%%%%%%%%%%%%%%%%%%%%%%%%%%%%%%%%%%%%%%%
It is well-established that the hyperfine interaction 
between quarks in a baryon, 
mediated by an effective one-gluon-exchange  (OGE) interaction, explains 
the major part of the baryon octet and decuplet mass difference --  
e.g. the nucleon-$\Delta$ and $\Sigma - \Lambda$ mass 
differences~\cite{De Rujula:1975ge,Chodos:1974pn}. 
This spin-spin  (OGE) interaction will therefore also contribute 
when an external probe interacts with the three-quark baryon state. 
That is, the probe not only senses a single quark current 
but a two-quark current as well. 
This two-quark current has an intermediate quark propagator connecting the 
probe and the hyperfine-interaction vertices as illustrated in Fig.~1. 

%\vspace{5mm} 

In the exploration of the two-quark axial current by 
H\o gaasen and Myhrer~\cite{Hogaasen:1988jd}, the MIT bag model 
was used and the confined quark propagator was 
written as a sum over quark eigenmodes.  
The dominant OGE exchange current corrections to the octet
magnetic moments and  semi-leptonic decays were found to come from the 
intermediate p-wave anti-quark states.  
They found that this correction was vital in reproducing 
not only the observed 
magnetic moment ratio of the two 
baryons $\Xi^-$ to $\Lambda$~\footnote{ 
According to PDG this ratio is larger than one; 
(-0.6507 $\pm$ 0.0025 n.m.) /  (-0.613 $\pm$ 0.004 n.m.).  
Most models without the OGE find this ratio smaller than one. 
}
but also the unusual strength of the decay 
$\Sigma^- \to n + e^- + \bar{\nu}_e$. 

%\vspace{5mm}

Myhrer and Thomas~\cite{Myhrer:1988ap} realized the importance of this  
OGE correction to the flavor singlet axial 
charge and hence to $\Sigma_{quark}$ in 
Eq.(\ref{eq:Bjorken1}). They concluded that this OGE correction reduced the 
fraction of the spin carried by the quarks in the proton by 0.15, i.e. 
$\Sigma_{quark} \to \Sigma_{quark} - 3 G$. 
The correction term $G \simeq 0.05$, found by 
summing over the quark states, is proportional to 
$\alpha_s$ times a bag model matrix element~~\cite{Hogaasen:1988jd} where 
$\alpha_s$ is determined by the ``bare" nucleon - $\Delta$ mass difference.  
Again the spin is lost to angular momentum of quarks and antiquarks, 
the latter predominantly in p-waves. 

The importance of the hyperfine interaction among quarks with respect 
to the large-$x$ behaviour of parton distribution functions 
was discussed by  
Close and Thomas~\cite{Close:1988br} and Isgur~\cite{Isgur:1999}.

%\vspace{5mm}

%%%%%%%%%%%%%%%%%%%%%%%%%%%%%%%%%%%%%%%%%%%%%%%%%%%%%%%%%%%%%%%%%%
\begin{center}
{\bf 3.3 The pion cloud } 
\end{center}
%%%%%%%%%%%%%%%%%%%%%%%%%%%%%%%%%%%%%%%%%%%%%%%%%%%%%%%%%%%%%%%%%%%

The virtual pion emission and absorption by the nucleon quark core is 
an effective implementation of the chiral symmetry requirements and 
is crucial in describing the charge radii of the 
proton and neutron, as well as other 
properties of the 
nucleon~\cite{Theberge:1980ye,Thomas:1982kv,Tsushima:1988,mbx:1981}. 
The cloudy bag model (CBM)~\cite{Theberge:1980ye,Thomas:1982kv} 
incorporates the 
chiral symmetry requirements and is highly successful 
in describing baryon properties.
In this model the nucleon consists of a bare nucleon, $| N >$,
with a probability $Z \sim 1 - P_{N \pi } - P_{\Delta \pi} \sim 0.7$,
in addition to being described as a nucleon $N$ and a pion 
and a $\Delta$ and a pion, 
with probabilities $P_{N \pi} \sim 0.20 - 0.25$ 
and $P_{\Delta \pi} \sim 0.05 - 0.10$, respectively.
The phenomenological constraints on these probablilities were discussed in 
Refs.~\cite{Speth:1996pz,Melnitchouk:1998rv,Thomas:2007}.
The most well-known of these constraints is 
associated with the excess of $\bar{d}$ 
over $\bar{u}$ quarks in the proton, 
predicted on the basis of the CBM~\cite{Thomas:1983fh}. 
Indeed, to first order the integral of $\bar{d}(x) - \bar{u}(x)$ is 
$2/3P_{N\pi}-P_{\Delta \pi}/3$, which is consistent with the experimental 
data~\cite{Arneodo:1996kd} if $P_{N\pi}$ and $P_{\Delta \pi}$ 
lie within the ranges just quoted. 

%\vspace{5mm} 

The effect of the pion cloud  on the quark 
spin contribution was investigated early by 
Schreiber and Thomas~\cite{Schreiber:1988uw}. They wrote the corrections to the spin sum-rules for the proton and neutron explicitly in terms of the probabilities set out above. 
{}For the present purposes it is helpful to 
rewrite the results of Ref.~\cite{Schreiber:1988uw}.  
If we consider the flavor singlet combination 
the pion cloud correction modifies the quark 
spin contribution in the following manner:
\begin{equation}
\Sigma_{quark} \rightarrow  \left(Z - \frac{1}{3} P_{N \pi} +\frac{5}{3} P_{\Delta \pi} \right) 
\Sigma_{quark} \, .
\label{eq:pion}
\end{equation}
The critical feature of the pion cloud correction in Eq.~(\ref{eq:pion}) 
is that the Clebsch-Gordon algebra for coupling the spin of the proton and 
the orbital angular momentum of the pion in the 
$N \pi$ Fock state favors a spin down nucleon and a 
pion with +1 unit of orbital angular momentum. 
This too has the effect of replacing quark spin by 
quark and anti-quark orbital angular momentum. 
Note that in the $\Delta \pi$ Fock component the 
spin of the baryon tends to point up (and the pion angular momentum down), 
thus enhancing the quark spin. 
Nevertheless, the wave function renormalization factor, $Z$, dominates, 
yielding a reduction by a factor between 0.7 and 0.8 for 
the range of probabilities quoted above.  

%\vspace{5mm} 

%%%%%%%%%%%%%%%%%%%%%%%%%%%%%%%%%%%%%%%%%%%%%%%%%%%%%%%%%%%%%%%%%%
\begin{center}
{\bf 3.4 The possible polarized gluon contribution $\Delta G$ } 
\end{center}
%%%%%%%%%%%%%%%%%%%%%%%%%%%%%%%%%%%%%%%%%%%%%%%%%%%%%%%%%%%%%%%%%%%

%\vspace{3mm} 

We have used a model of confined quarks 
to compute the matrix elements of the axial current to find 
$\Sigma$ and $g_A$ values relevant in the limit 
$Q^2 \to \infty$. 
Our model result, $\Sigma_{quark} \in (0.35,0.40)$~\cite{mt08}, 
agrees very well with the experimental value 
$\Sigma $ -- c.f. Eq.(\ref{eq:SigmaEXP}). 
Although there is no unambiguous way to identify the scale associated with 
the chiral quark model, many authors (see e.g., 
Refs.~\cite{Parisi:1973nx,Signal:1988vf,Gluck:1988ey}) 
have  made the  
observation that a valence dominated quark model can only match 
experiment for 
parton distribution functions at a relatively low $Q^2$ scale. 
We argue that the quark model evaluation of the proton 
matrix element of the flavor singlet axial current 
corresponds to the first term in 
Eq.(\ref{eq:UA1}). 
To explain the reasoning in more detail we observe the following. 
In the limit $Q^2 \to \infty$ the product $\alpha_s(Q^2)\Delta G(Q^2)$ 
in Eq.(\ref{eq:UA1}) tends to a constant~\cite{ccm88}. 
In the same $Q^2$ limit the flavor singlet part of the $x$-integral over 
$g_1(x,Q^2)$ of Eq.(\ref{eq:Bjorken1}) equals the low-energy
proton matrix element of the singlet axial current 
\begin{eqnarray}
\sum_{i = u, d, s} \ < p | \bar{q}_i \gamma^\mu \gamma_5 q_i | p > \; , 
\end{eqnarray} 
which contains a contribution from the  quarks as well as 
the axial anomaly term involving 
$\alpha_s(Q^2)\Delta G(Q^2)$. 
The separation of the two contributions in Eq.(\ref{eq:UA1}) 
is most naturally made in terms of the range of integration over 
the transverse quark momenta,  $k_\perp^2$.  
The contribution of the axial anomaly in Eq.(\ref{eq:UA1}) 
relies on the scale separations~\cite{ccm88}: 
$m_{quark}^2 \ll (1/R_{confinement}^2) \sim  p_{gluon}^2 \ll Q^2$,  
and the integration over transverse momenta is dominated by the 
range $ k_\perp^2 \ge Q^2$~\cite{ccm88,Bass:1991yx}. For comparison, 
the contribution from valence quarks is dominated by 
relatively low transverse momenta, 
$ k_\perp^2 < 1$(GeV/c)$^2 <Q^2$. 
The valence parton distributions calculated in this way 
correspond to the original Gribov-Lipatov definition of a parton distribution 
at scale $Q^2$. 
Numerically, we note that if the value of 
$\Delta G \simeq 0.4$ at $Q^2 \simeq 3$(GeV/c)$^2$ as inferred from recent 
results at RHIC~\cite{:2008px}, then the second term in Eq.(\ref{eq:UA1}) 
is estimated to contribute about 0.06 to $\Sigma$, using 
$\alpha_s(Q^2)\simeq 0.3$ at this $Q^2$ scale.

%%%%%%%%%%%%%%%%%%%%%%%%%%%%%%%%%%%%%%%%%%%%%%%%%%%%%%%%%%%%%%%%%%%%%%%%%
\begin{center}
{\bf 4. Lattice QCD calculations of quark orbital angular momentum }
\end{center}
%%%%%%%%%%%%%%%%%%%%%%%%%%%%%%%%%%%%%%%%%%%%%%%%%%%%%%%%%%%%%%%%%%%%%%%%%%% 

The results of the last section imply that a large fraction of the proton spin 
is carried by quark and anti-quark orbital angular momenta. The distributions of
the orbital angular momenta for $u + \bar{u}$ and $d+\bar{d}$ 
were extracted from our model results by 
Thomas~\cite{Thomas:2008} and he found that 
$L^{u+\bar{u}} \simeq .50$ and 
$L^{d+\bar{d}} \simeq 0.12$, which give  
$J^{u+\bar{u}} = L^{u+\bar{u}} +\Delta u/2 \simeq 0.72$   and 
$J^{d+\bar{d}} = L^{d+\bar{d}} +\Delta d/2 \simeq - 0.10$. 
These values should be compared to the QCD lattice results for 
orbital angular momentum as evaluated by the LHPC collaboration~\cite{LHPC08}. 
The lattice QCD $ L^{u+\bar{u}}$ and $ L^{d+\bar{d}}$ values  
were however evaluated at a 
scale of about 4 GeV$^2$, much larger than 
the scale relevant for the chiral quark model. 
In order to
compare the angular momentum results of the model with the lattice QCD results, 
Thomas~\cite{Thomas:2008} used QCD evolution equations for angular momentum as 
outlined in the work by Ji and collaborators~\cite{jth96}.~\footnote{ 
Ji~\cite{Ji1997} introduced a gauge-invariant decomposition of the nucleon spin
into quark helicity, quark orbital and gluon contributions.
Recently there has been some debate on 
the proper way to define the quark orbital angular momentum, see e.g. 
Refs.~\cite{Chen:2008ag,Ji:2008uh,Chen:2008ja,Chen:2009mr,Burkardt:2008ua}. 
}
Thomas showed that  the QCD evolution has a dramatic influence on 
the different quarks' orbital angular momenta as a function of $Q^2$ and that 
the model results, which are derived based on the results of 
the previous section, are consistent with the lattice QCD results 
and also with results for $J^{u+\bar{u}}$ and $J^{d+\bar{d}}$, which 
have been extracted from DVCS experiments at DESY~\cite{HERMES:2006} 
and JLab~\cite{JLAB:2007} using the model of 
Goeke {\it et al.}~\cite{Goeke:2001} -- see also~\cite{vgg:1999}. 

Although the predictions of the Myhrer-Thomas (MT) work are consistent with both 
the lattice QCD data and the model dependent analysis of recent experiments, 
it is important to realize that at present this is not a compelling statement. 
The nature of the QCD evolution is such that predictions which differ 
dramatically at the model scale tend to be much closer at 4 GeV$^2$. 
{}For example, a change in the MT value for $L^u - L^d$ 
by 0.10 at the model scale 
leads to a change of only 0.01 at 4 GeV$^2$. Clearly, this makes the 
challenge of making measurements, either experimentally or on the lattice, 
which clearly discriminate between models, very difficult indeed.
{}Furthermore, we should note that the current lattice 
calculations suffer from the omission of ``disconnected quark loops'', which 
contain the effect of the axial anomaly and this introduces a completely 
unknown systematic error -- quite apart from the usual uncertainties of 
taking the chiral limit, the lattice spacing to zero and the volume to 
infinity.

It is also very important to note that, as pointed out by Wakamatsu 
and Tsujimoto, whereas the chiral quark soliton model usually yields 
very similar results to those found in the CBM, in the case of 
$L^u - L^d$ they are completely 
different~\cite{Wakamatsu:2005vk}. Indeed, 
the non-linear pion fields 
in the chiral quark soliton model yield a large negative prediction for 
this quantity at the model scale. For the moment the uncertainties that 
we summarised in the previous paragraph make discrimination between the 
models on this basis impossible at present but this may prove a critical 
discriminator as the lattice computations improve over the next few years.
 
%\vspace{5mm}

%%%%%%%%%%%%%%%%%%%%%%%%%%%%%%%%%%%%%%%%%%%%%%%%%%%%%%%%%%%%%%%%%%%%%%%%%
\begin{center}
{\bf 5. The GPDF and quark angular momentum }
%%%%%%%%%%%%%%%%%%%%%%%%%%%%%%%%%%%%%%%%%%%%%%%%%%%%%%%%%%%%%%%%%%%%%%%%% 
%
%%%%%%%%%%%%%%%%%%%%%%%%%%%%%%%%%%%%%%%%%%%%%%%%%%%%%%%%%%%%%%%%%%%%%%%%% 
\end{center} 

Some time ago 
the transverse quark degrees of freedom in hadrons were explored 
in order to gain some understanding of  
the high $Q^2$ large scattering angle behavior of 
$pp$ cross section, analyzing power and $A_{NN}$.  
In these large $Q^2$ measurements the short-range (``hard") scattering
processes dominate. 
However, the transverse hadronic dimensions, 
which are determined by confinement and generate  
``medium -range" interactions among the valence quarks, 
can interfere and successfully explain these  
$Q^2$ phenomena~\cite{Ralston86,cmm92}.  
In addition, these medium-range interactions gave some possible insight as to 
why the measured  exclusive cross sections not only exhibit 
the expected hard-scattering 
$Q^2$ scaling behavior but also explain the superimposed 
oscillatory  
behavior of the cross sections versus  $Q^2$ 
(for a fixed large scattering angle) 
in measured exclusive hadronic reactions, e.g. pp and 
$\pi$p scattering. 
The inferred conclusion was that even if helicity is conserved in the 
perturbative quark-gluon processes, helicity is not necessarily 
conserved at the hadronic level. 
In short, the transverse components of a hadron introduce 
quark angular momentum and 
possible quark spin-orbit interactions 
into the description of hadronic spin high $Q^2$ observables 
at large scattering angles.

\vspace{5mm}

To examine the importance of the quark orbital 
angular momentum one has to study the 
Generalized Parton Correlation Functions (GPCF) which 
could be extracted from measurements. 
A concise overview of  possible measurements of GPCF and their connection to 
the quarks transverse degrees of freedom, has been presented in a recent 
COMPASS report~\cite{compass09}. 
A detailed exposition of recently proposed non-trivial relations between 
generalized and transverse momentum dependent parton distributions 
(GPD and TMD, respectively) 
are given in Ref.~\cite{Meissner:2009}. 
According to Ref.~\cite{Meissner:2009} there are no model-independent  
relations between GPD and TMD functions. 
It is unfortunate that model considerations are
necessary in order to extract the transverse momentum dependences of GPCF from 
measured observables. 
Recently Efremov {\it et al.}~\cite{Efremov:2009} used a 
covariant model of {\it free} quarks to discuss relations between  
GPDs and TMDs which can be derived from GPCF.
Some of these relations have also 
been derived in several other parton model studies, 
see e.g., Ref.~\cite{Efremov:2009} and references therein. 
As argued by Burkardt~\cite{Burkardt:2009}, 
parton distributions in impact parameter space show significant 
deviation from axial symmetry when, for example, 
the proton is transversely polarized. 
Extracting this axial asymmetry using, for example, the Sivers function  
$f_{1T}^{\perp \ q} (x, Q^2)$ could provide useful information about 
quark angular momentum, however as mentioned   
one has to resort to a model to extract the necessary information.  
Ma {\it et al.}~\cite{Ma:1998} showed that 
pretzelosity, another TMD, could give useful information about the  
quark angular momentum but a quark-diquark model is necessary to extract 
this information. 
A recent model discussion on pretzelosity can be found in 
Avakian {\it et al.}~\cite{Avakian:2008mu}.
According to Meissner {\it et al.}~\cite{Meissner:2009} the TMDs 
are determined when hadron momentum transfer $\Delta = p^\prime - p = 0$ 
after an integration over $k^-$ of the GPCFs.
In the forward limit, $\Delta \simeq 0$,  
a word of caution has been advocated by Szczepaniak 
{\it et al.}~\cite{Szczepaniak:2006is,Brodsky:2009bp} who show that 
it is a non-trivial task to  extract TMDs from measured data. 
They also present evidence for a fixed $J=0$ pole contribution to deep 
virtual Compton scattering at all momentum transfers. This acts like 
a subtraction in the dispersion relation for the 
Compton amplitude~\cite{Brodsky:2008qu,Brodsky:2009bp}. 
%%%%%%%%%%%%%%%%%%%%%%%%%%%%%%%%%%%%%%%%%%%%%%%%%%%%%%%%%%%%%%%%%%%%%%%%%%%%%%%
%%%%%%%%%
%The concept of positivity appears to be extremely useful to constrain relations among 
%spin observables~\cite{Soffer95}, see the review~\cite{Artru:2009} and a recent 
%discussion by Bourrely {\it et al.}~\cite{bbs09}.
%%%%%%%%%
%%%%%%%%% 
%%%%%%%%%%%%%%%%%%%%%%%%%%%%%%%%%%%%%%%%%%%%%%%%%%%%%%%%%%%%%%%%%%%%%%%%%%%%%
%%%%%%%%%%%%%%%%%%%%%%%%%%%%%%%%%%%%%%%%%%%%%%%%%%%%%%%%%%%%%%%%%%%%%%%%%%%%% 
%\vspace{5mm}
%
%Elliot Leader has one other preprint~\cite{Leader2008}, 
%which is worth reading and maybe fits into my section 5. 
%Part of the discussion in Ref.~\cite{Leader2008} 
%centers on the correct way to take the expectation values of 
%the transverse angular momentum operator. See here also the footnote on page 10.  
%
%\vspace{5mm} 
%
%{\bf Tony. here are two references I don't know if they are relevant or not. 
%We can drop both.}: \\ 
%W.-D. Nowak, ``The angular momentum structure of the nucleon", 
%arXiv:hep-ph/0812.2679 (2008). \\
%U. D'Alesio, ``Exploring the transverse spin structure of the nucleon", 
%arXiv:hep-ph/0809.3162 (2008). 
%%%%%%%%%%%%%%%%%%%%%%%%%%%%%%%%%%%%%%%%%%%%%%%%%%%%%%%%%%%%%%%%%%%%%%%%%%%%%%%

%%%%%%%%%%%%%%%%%%%%%%%%%%%%%%%%%%%%%%%%%%%%%%%%%%%%%%%%%%%%%%%%%%%%%%%%%
\begin{center}
{\bf 6. Concluding remarks }
\end{center}
%%%%%%%%%%%%%%%%%%%%%%%%%%%%%%%%%%%%%%%%%%%%%%%%%%%%%%%%%%%%%%%%%%%%%%%%% 

We have seen that the latest data on the proton spin sum-rule, which 
yields the result that the fraction of the spin of the proton carried by 
its quarks, $\Sigma = 33 \pm 3 \pm 5$\%, is very naturally explained within a 
relativistic quark model that includes the effective 
one-gluon-exchange hyperfine 
interaction and respects chiral symmetry. In many ways these are what might 
be regarded as the basic ingredients of a modern model of nucleon structure 
and it is very satisfying that the proton spin crisis, which has caused such 
consternation in the nuclear and particle physics communities over the past 20 
years, can be explained this way. 

The role of the axial anomaly is now known to be considerably smaller than 
was once hoped, with $\Delta G$ now most likely between 0 and 0.4~\cite{:2008px} 
at $Q^2$  of order 4 GeV$^2$. 
Nevertheless, even such a small contribution will 
become significant as the precision with which $\Sigma$ is determined 
increases. Indeed, the corresponding contribution 
to $\Sigma$ from gluons in this 
range would be between -0.06 and 0. As a consequence of this, the gauge 
invariant quark spins, $\Delta u, \Delta d$ and $\Delta s$ would each 
receive a gluonic contribution as large as -0.02. A simple kaon loop 
calculation suggests a chiral contribution to $\Delta s$ of order 
-0.01, resulting in a total value of $\Delta s$ as large as -0.03. 
Testing this directly, for example through neutral current neutrino-proton 
elastic scattering, would be extremely valuable. 

It is interesting to note the upper limit on the 
magnitude of $\Delta s$ derived in this way, namely $\Delta s = -0.03$,  
is considerably lower 
than the value derived from $\Sigma = \Delta u + \Delta d 
+ \Delta s$ (given above) and the value of $g_A^8 = \Delta u + \Delta d 
- 2 \Delta s = 0.57 \pm 0.03$ usually derived from hyperon $\beta$-decay, 
namely $\Delta s = -0.08 \pm 0.01 \pm 0.02$. Of course, it has been 
argued~\cite{Jaffe:1989jz} 
that the error in applying SU(3) symmetry to the octet axial charges 
may be as large as 20\%, which could bring the value of $\Delta s$ derived 
from $\Sigma$ and $g_A^8$ as low as $-0.04 \pm 0.01 \pm 0.02$, which 
would be in much better agreement. It is clearly important to carefully 
derive theoretical values for {\it both} $\Sigma$ and $g_A^8$ within 
any model applied to the spin problem.  

While Occham's razor suggests that the simplest explanation is almost certainly 
the correct one, we note that there are several other proposals which are 
also able to explain the experimental result. Using a generalization of the 
Goldberger-Treiman relation to the singlet case,  
Shore and Veneziano~\cite{Shore:1990zu,Narison:1994hv,Narison:1998aq} 
predicted a reduction of about 50\% in the naive relativistic expectation for 
$\Sigma$. This model has interesting predictions for the corresponding 
suppression in other hadrons. 

Another fascinating suggestion from 
Bass~\cite{Bass:2004xa,Bass:2003vp,Bass:2000kr} 
sees much of the spin of the proton
tied up in the topological structure of the gluon fields. This would show 
up only through a $J=1$ fixed pole which contributes to the structure 
functions as a $\delta$-function at $x=0$. One signature of such an effect 
would be a difference in the value of $\Sigma$ extracted from neutrino-proton 
elastic scattering from that obtained in deep inelastic scattering -- where 
the  $\delta$-function at $x=0$ is unmeasureable.   

Finally, we note that while the gross violation of quark model expectations 
has now been removed, so that in our view the ``spin crisis'' has been solved, 
the problem of understanding in detail how the spin of the proton is 
carried by its quarks and gluons is now of great interest. On the scale of 
one half, it matters a great deal whether $\Delta G$ is 0 or 0.4. The 
Myhrer-Thomas explanation of the spin crisis implies that much of the 
proton spin is carried as orbital angular momentum and it is critical 
to find ways to pin this down. We have seen that the study 
of GPDs through both lattice QCD and experiment will be crucial in this quest 
and the 12 GeV upgrade at Jefferson Lab~\cite{Thomas:2007zza} 
is ideally suited to play a key role, 
at least in the valence region. For the sea-quark region 
we may well need a high 
luminosity electron-ion collider~\cite{Thomas:2009ei} 
but that may take a little longer. 

%\vspace{5mm}

{\bf Acknowledgements}
This work was supported by the US Department Energy, 
Office of Nuclear Physics, through contract no. DE-AC05-06OR23177, 
under which Jefferson Science Associates operates Jefferson Lab 
and by the NSF grant PHY-0758114.

%%%%%%%%%%%%%%%%%%%%%%%%%%%%%%%%%%%%%%%%%%%%%%%%%%%%%%%%%%%%%%%%%%%%%%% 

%
\end{document}